\documentclass[a4paper, 10pt]{wlpeerj}
\usepackage{hyperref}
\usepackage{cleveref}
\usepackage{listings}
\title{\emph{flippy}: User friendly and open source framework for lipid membrane simulations}

\author[1]{George Dadunashvili}
\author[1]{Timon Idema}
\affil[1]{Department of Bionanoscience, Kavli Institute of Nanoscience, Delft University of Technology, The Netherlands}
\corrauthor[1]{Timon Idema}{t.idema@tudelft.nl}

\begin{abstract}
    Animal cells are both encapsulated and subdivided by lipid bilayer membranes. Beyond just acting as boundaries, these membranes' shapes influence the function of cells and their compartments. Physically, membranes are two-dimensional fluids with complex elastic behavior, which makes it impossible, for all but a few simple cases, to predict membrane shapes analytically. Instead, the shape and behavior of biological membranes can be determined by simulations. However, the setup and use of such simulations require a significant programming background. The availability of open-source and user-friendly packages for simulating biological membranes needs improvement. \\
    Here, we present \emph{flippy}, an open-source package for simulating lipid membrane shapes, their interaction with proteins or external particles, and the effect of external forces. Our goal is to provide a tool that is easy to use without sacrificing performance or versatility. \emph{flippy} is an implementation of a dynamically triangulated membrane. We use a precise yet fast algorithm for calculating the geometric properties of membranes and can also account for local spontaneous curvature, a feature not all discretizations allow.
    Finally, in \emph{flippy} we can also include regions of purely elastic (non-fluid) membranes and thus explore various shapes encountered in living systems.
\end{abstract}

\begin{document}

    \flushbottom
    \maketitle
    \thispagestyle{empty}

    \section*{Background}
    Lipid bilayer membranes form the envelopes of animal cells and of many organelles contained in them.
    These biological membranes are highly flexible materials, capable of adopting many nontrivial shapes corresponding to specific cell functions and responding to environmental circumstances.
    Therefore, we can infer which processes occur inside a cell or organelle from the shapes of their membranes~\cite{frey2021more}.
    Inversely, in synthetic biology, membranes can be manipulated to mimic growth and division processes of living cells~\cite{godino2019novo, litschel2018beating, steinkuhler2020controlled, rideau2018liposomes}.
    Achieving symmetric, stable division over many generations is a significant challenge in the bottom-up assembly of living cells.
    A crucial part of the problem is understanding how external mechanical and chemical cues drive membrane reshaping.
    Predicting the shapes of membranes analytically is very difficult and therefore limited to cases of membranes with few constraints and high symmetry.
    Even numeric solutions to analytic equations are usually only possible in highly symmetric cases.
    In order to predict membrane shapes for generic problems, we need to use simulations.
    Full atomistic simulations are out of the question due to computational constraints when we are interested in large-scale membrane reshaping.
    Luckily several types of simulations describe the membrane behavior on a large scale, like self-assembled membranes~\cite{yuan2010one}, phase fields based methods~\cite{biben2005phase}, and dynamically triangulated membrane Monte Carlo (DTMMC) simulations~\cite{gompper1997network}.
    The latter method is based on minimizing membrane surface energy, which makes interpreting results easy and leaves an opportunity to connect the findings to an analytical model~\cite{gueguen2017fluctuation}.
    Thus, it is not surprising that the method of dynamically triangulated membrane simulations has found broad adoption in the field and has been used to model diverse set of experimental systems from membranes responding to osmotic conditions~\cite{gueguen2017fluctuation} and shear flow~\cite{noguchi2004fluid}, to interactions of membranes with colloidal particles~\cite{vsaric2012mechanism, vsaric2012fluid,bahrami2012tubulation, van2016lipid, vahid2017curvature} and with proteins~\cite{helle2017mechanical, kumar2019tubulation, fovsnarivc2019theoretical}.

    \section*{Implementation}
    While DTMMC simulations are widely used, simulation codes are rarely published.
    We therefore wrote the \emph{flippy} software as an open-source package.
    While DTMMC simulations are popular, they are hard to write and even harder to optimize.
    Even with only basic functionality, a DTMMC code quickly becomes large and hard to maintain.
    Therefore, to make further progress in the development of DTMMC simulations, we need an open-source library with a vibrant community and developer base around it.
    We want our package to help biophysicists to get to implementing the specifics of their system without needing to reinvent the wheel and code the whole dynamically triangulated membrane from scratch.
    \emph{flippy} is designed with this objective in mind.\\
    An ideal simulation framework for membranes would not involve programming at all on the end user's side.
    Simulating a membrane under a specific physical constraint would be like conducting an experiment.
    A fully interactive framework would drastically reduce the barrier to simulations, only require understanding the experimental setup, and enable researchers to directly compare their results to simulations.
    However, this ideal case of an interactive framework is hard to implement in a vacuum.
    While keeping it in mind as an end-goal, we decided to start with a more manageable task.
    Even though using a \texttt{c++} library requires much more knowledge than just using interactive software, we keep user-friendliness and a high level of abstraction as our primary goals.
    The implementation of \emph{flippy} as a \texttt{c++} library instead of a domain-specific scripting language allows users the flexibility to incorporate it in existing code bases and easily extend it, thus contributing to our goals of community-based growth.
    Having an implementation in a compiled language additionally allows the users to create fast simulations, increasing the range of systems that can be simulated by \emph{flippy} in a reasonable time.\\
    The fact that \texttt{c++} does not have a centralized package manager usually makes it hard to obtain or use external libraries.
    In our experience, this is the most significant inconvenience related to using the \texttt{c++} language.
    To minimize this friction as much as possible, we opted to implement a header-only library and eliminate almost all external dependencies.
    Our code only relies on an external \textsc{JSON} parser to easily save simulation data.
    This parser is itself licensed under the same open source license as \emph{flippy}, which enabled us to bundle it with \emph{flippy}~\cite{Lohmann_JSON_for_Modern_2022}.
    This independence from external dependencies makes using our software package as easy as it gets for \texttt{c++} libraries.
    \subsection*{Code quality control}
    Every large code base is prone to hidden bugs and unexpected behaviors in new use cases.
    To minimize errors, we implemented an extensive unit testing framework, and we are happy to report that our code base has over $95\%$ coverage.
    Thus, almost every function implemented in our base is covered by at least one test case, and we intend our library of unit tests to grow continuously.
    We are aware that unit tests cannot guarantee that the code is bug-free, and we intend to use the bug reporting facilities of the GitHub repository to enable our users to report bugs and help improve the package.
    \subsection*{Mathematical details of the implementation}
    Since we aspire to a user-friendly framework, \emph{flippy} must implement commonly needed utilities that almost every DTMMC simulation will require.
    The triangulation provided by \emph{flippy} needs to do proper bookkeeping of several important geometric quantities, during the update of the triangulation, like the local curvature vector, local area, and local unit bending energy of each node.
    We also keep track of the global counterparts of these quantities, i.e.,\ the total area and total unit bending energy of the triangulated shape.
    These quantities are defined on continuous shapes, requiring a mathematically rigorous discretization on a triangulated lattice.
    By this, we mean that the discretized quantities should converge to their continuous counterparts for finer triangulations, and the simulation should become more precise with an increasing number of triangles.
    From all the above, the local curvature is most challenging to discretize and can lead to triangulation-dependent curvature energies, as demonstrated by Gompper and Kroll~\cite{gompper1996random}.
    We use an extension of the method proposed in~\cite{gompper1996random} for calculating local mean curvature.
    This extension was introduced by Meyer et al.~\cite{meyer2003discrete} to calculate the local area associated with a node more precisely and sidestep numerical problems that occur for triangulations containing obtuse triangles.
    Finally, we use the same expression for the node-associated volume as Guegen et al.\@\cite{gueguen2017fluctuation},
    which is fast to calculate since it only relies on already computed quantities.
    However, this node-associated volume does not have a physical meaning, it only sums to the correct total volume enveloped by a closed triangulation.

    \section*{Results}\label{sec:results}
\begin{figure}[!h]
        \includegraphics[width=\linewidth]{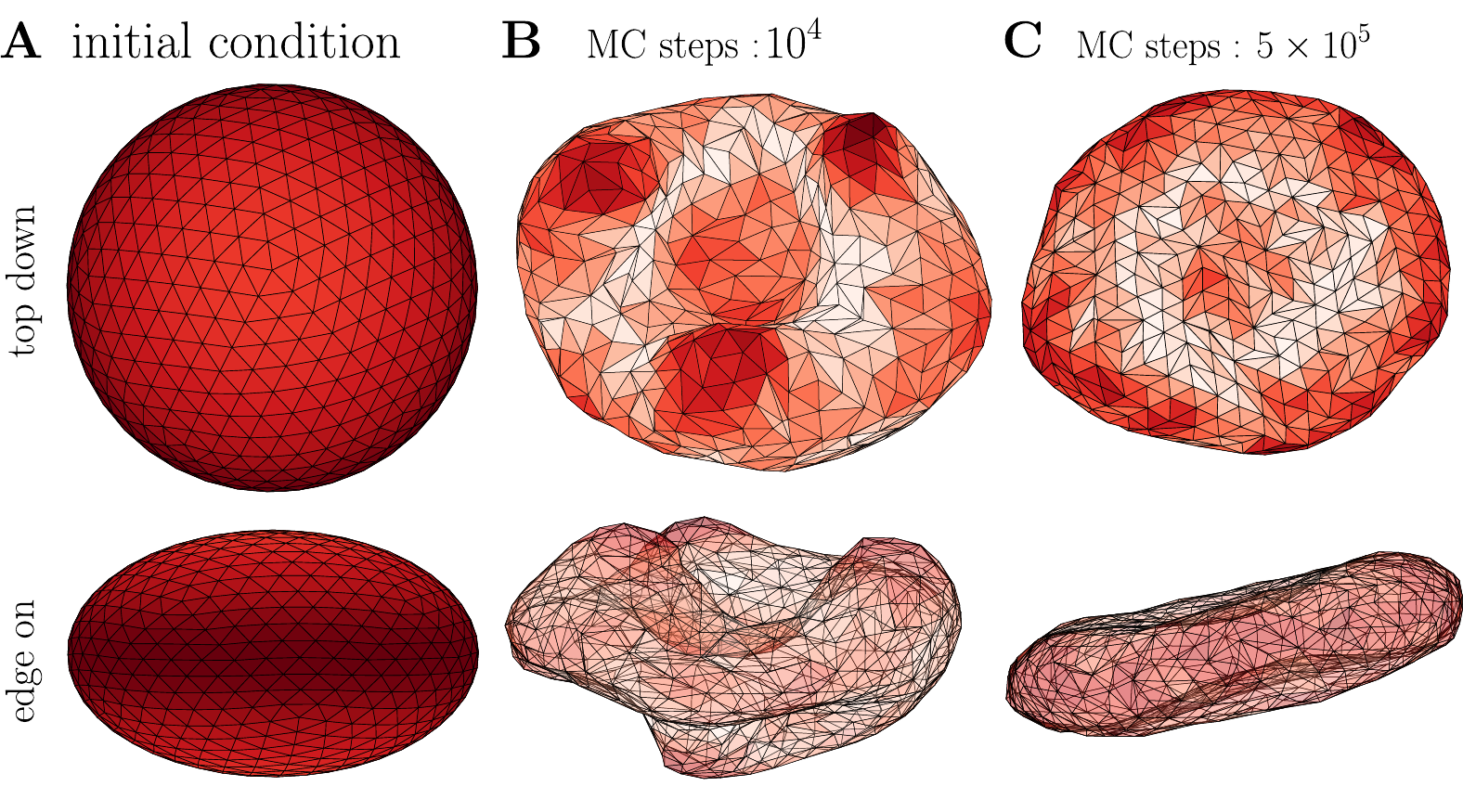}
        \caption{\textbf{Result of the Monte carlo Simulation} The triangulation consisted of $642$ nodes. \textbf{A}: Initial condition. A slightly oblate spheroid. \textbf{B}:  Simulation that ran for $10^{4}$ Monte Carlo steps per node (took 10 seconds). \textbf{C}: Simulation that ran for $5\times10^{5}$ Monte Carlo steps per node (took approximately $7$ minutes). C is not a longer run of B but a new simulation that ran longer than B. All simulations were done on Intel i7-8650U 1.9GHz processor.}
        \label{fig:RBC_series}
    \end{figure}

    In this section, we want to demonstrate \emph{flippy's} ability to abstract away the implementation details of a dynamic triangulation and Monte Carlo updating scheme.
    To this end, we go through the process of simulating a simple experimental system of a deflated giant unilamellar vesicle (GUV) and use \emph{flippy} to predict the equilibrium shape of the vesicle.
    In the following, we will only present the key elements of the code.
    The complete version of this simulation is provided on GitHub; for more details, please see the \emph{Summary and outlook} section.\\
    The system of a deflated GUV can be modeled by the following surface energy
    \begin{equation}\label{eq:surface_energy}
    E_{\mathrm{surf}}=\frac{\kappa}{2}\int \mathrm{d}A (2H)^2 + K_A\frac{(A-A_t)^2}{A_t} + K_V\frac{(V-V_t)^2}{V_t},
    \end{equation}
    where $\kappa$ is the bending rigidity and $H$ is the local mean curvature of the membrane.
    The integral $\int \mathrm{d}A$ ranges over the surface area of the vesicle.
    This part of the energy describes the tendency of the biological membranes to minimize their local square mean curvature~\cite{canham1970minimum, helfrich1973elastic}.
    The Lagrange multipliers $K_A$ and $K_V$ fix the area $A$ and volume $V$ to their target values $A_t=A_0=4\pi R_0^2$ and $V_t=0.6V_0=0.6\frac{4\pi}{3}R_0^3$, where $R_0$ is the radius of the initial (pre deflation) spherical GUV\@.
    Since the deflation of the vesicle does not change its area, we keep it fixed to the initial value.
    However, the target value of the volume is fixed to $60\%$ of the initial volume to account for deflation.
    We picked $60\%$ of the initial volume because, for this value, we expect the equilibrium configuration to be a biconcave shape, providing an easy visual way to judge the success of the simulation.
    However, the prediction of a biconcave shape is only precise for zero temperature, i.e., for $k_{B} T=0$.
    For simplicity, the following example describes a simulation performed at $k_{B} T=1$.
    Thus, the resulting shapes will be noisy and not perfectly biconcave.\\
    All the complexity of creating and maintaining a dynamic triangulation is hidden in three conceptual steps;
    define the energy, initiate a triangulation, and initiate an updater that will use the energy to update the triangulation.
    We can start with the definition of the energy function that implements the~\cref{eq:surface_energy}.
    Since the \texttt{MonteCarloUpdater} will use this energy, its signature needs to follow a specific convention that the updater will recognize,
    \begin{lstlisting}[language=C++,label={lst:surface_energy_signature}]
double surface_energy(fp::Node<double, unsigned int> const&,
                      fp::Triangulation<double, unsigned int> const& ,
                      EnergyParameters const& )
    \end{lstlisting}
    where the first argument needs to be a \emph{flippy} \texttt{Node} type, representing the node that is being updated, and the second argument needs to be a \texttt{Triangulation} type representing the triangulation that is being updated. The third argument can be any type and is intended to be a user-defined data struct containing all the parameters of the energy function.
    The actual function body is then a straightforward implementation of~\cref{eq:surface_energy}:
    \begin{lstlisting}[language=C++,label={lst:surface_energy}]
double surface_energy([[maybe_unused]]fp::Node<double, unsigned int> const& node,
                      fp::Triangulation<double, unsigned int> const& trg,
                      EnergyParameters const& prms){
    double V = trg.global_geometry().volume;
    double A = trg.global_geometry().area;
    double dV = V-prms.V_t;
    double dA = A-prms.A_t;
    double energy = prms.kappa*trg.global_geometry().unit_bending_energy +
                    prms.K_V*dV*dV/prms.V_t + prms.K_A*dA*dA/prms.A_t;
    return energy;
}
    \end{lstlisting}
    Here the first variable in the function signature is designated \texttt{[[maybe\_unused]]} since in this particular implementation of the energy, we are not interested in the local properties of any given node and thus do not use this variable.\\
    The second step in the implementation of the model is to declare a triangulation:
    \begin{lstlisting}[language=C++,label={lst:Tr_declaration}]
fp::Triangulation<double, unsigned int> tr(n_triang, R_0, r_Verlet);
    \end{lstlisting}
    where the template parameters \texttt{double} and \texttt{unsigned int} specify which internal representation of floating point and integer numbers the \texttt{Triangulation} class is supposed to use.
    The first argument of the instantiation \texttt{n\_triang} specifies the level of triangulation, which sets the fineness of the mesh.
    The second argument, \texttt{R\_0}, sets the initial radius of the triangulated sphere, and the last argument \texttt{r\_Verlet}, relates to the implementation of membrane self-intersection avoidance.
    \emph{flippy} implements a Verlet list to check spatial closeness of the nodes efficiently~\cite{verlet1967computer}.\\
    The third step is to declare a Monte Carlo updater that will use the energy function to update the triangulation according to a Metropolis algorithm~\cite{metropolis1953equation}:
    \begin{lstlisting}[language=C++,label={lst:mcu_declaration}]
fp::MonteCarloUpdater<double, unsigned int, EnergyParameters,
                      std::mt19937, fp::SPHERICAL_TRIANGULATION>
                     mc_updater(tr, prms, surface_energy, rng, l_min, l_max);

    \end{lstlisting}
    The signature of this class instantiation is quite large since the updater needs to know the energy function, all necessary update parameters, and the triangulation.
    The first two template parameters specify the internal representation of numbers, just like in the case of the \texttt{Triangulation} class.
    These parameters must be the same in both cases.
    \texttt{EnergyParameters} specifies the user-defined struct type name that contains the parameters used inside the energy function.
    \texttt{std::mt19937} specifies the type of the random number generator that we will provide to the updater for generating random numbers for the Metropolis algorithm.
    The last parameter specifies the triangulation type (currently, spherical and planar triangulations are possible).
    The instance of the updater itself has six arguments.
    The first four provide the updater with references to the already declared instances of triangulation class, energy parameters struct, energy function, and a random number generator.
    The last two arguments specify minimum and maximum allowed distances between the triangulation nodes.\\
    All that is left to do is to create an update loop that specifies in what order and how often we want to update the triangulation.
    \begin{lstlisting}[language=C++,label={lst:update_loop}]
for(unsigned int mc_step=0; mc_step<max_mc_steps; ++mc_step){
    for (unsigned int node_id: shuffled_ids) {
        displ = {displ_distr(rng), displ_distr(rng), displ_distr(rng)};
        mc_updater.move_MC_updater(guv[node_id], displ);
    }
    std::shuffle(shuffled_ids.begin(), shuffled_ids.end(), rng);
    for (unsigned int node_id: shuffled_ids) {
        mc_updater.flip_MC_updater(guv[node_id]);
    }
}
    \end{lstlisting}
    where in every update step, we loop over each node and use the methods of the \texttt{MonteCarloUpdater} class to move nodes and flip bonds.
    Between the loops where we \emph{move} and \emph{flip} the nodes, we shuffle the \texttt{shuffled\_ids} vector, which was defined before the loop and contains the ids of the nodes.
    The shuffling step ensures that we iterate randomly through the nodes at each Monte Carlo step and do not introduce unwanted correlations between node updates.
    This loop represents the logic of the experiment that we want to model.
    In this case, the experiment is simple; we are equilibrating a vesicle at a constant temperature (starting from slightly unphysical initial conditions of mismatching volume).
    This loop will become more complex as the needs of the simulation will grow.
    However, this corresponds to the true complexity that arises from the system itself and not from the implementation.
    Some higher stages of complexity might require a more sophisticated updater.
    The \texttt{MonteCarloUpdater} class is provided by flippy because a Metropolis updating scheme is a popular one.
    However, the \texttt{Triangulation} class itself is completely agnostic towards the updating scheme that is used on it.
    The user is free to implement another updating scheme if the Metropolis algorithm is not suitable to their problem and still be able to use the triangulation provided by flippy.
    This also enables us to easily extend flippy with new updaters.\\
    Finally, to make saving the state of the simulation easy, \emph{flippy}'s \texttt{Triangulation} class has a method that saves the representation of the data as a \emph{JSON} object, which is a text-based human-readable data format~\cite{ecma2017standard}.
    \emph{flippy} comes bundled with an open source \emph{JSON} parser~\cite{Lohmann_JSON_for_Modern_2022}.
    A single statement is sufficient to create \emph{JSON} data of the current state of the triangulation
    \begin{lstlisting}[language=C++,label={lst:make_data}]
    fp::Json data = tr.make_egg_data();
    \end{lstlisting}
    and a utility function in flippy allows the saving of this data to a text file as follows:
    \begin{lstlisting}[language=C++,label={lst:save_data}]
    fp::json_dump("test_run_final", data);
    \end{lstlisting}
    The \texttt{make\_egg\_data} methods naming refers to the fact that this \emph{JSON} data contains the necessary information to reinitialize the triangulation (like an egg contains all the nutrients for the chicken that will hatch from it), thus allowing the user to continue simulation from a save-file.
    If we use this simple code~\cite{Dadunashvili_flippy_demo_2023} (which is comfortably below $100$ lines, including all imports, variable definitions, and comments), we will obtain (in a few minutes) the expected biconcave shape (see \cref{fig:RBC_series} B and C).
    This example clearly shows that \emph{flippy} is capable of simulating a simple physical system in few lines of code, where all unnecessary complexity is abstracted away in the library and all the complexity that is still left in the user written code, contains necessary information about specific characteristics of the simulated system in question.
    Importantly, this abstraction and simplicity don't come at the cost of unreasonable runtime of the simulation.

    \section*{Summary and outlook}
    The source code of \emph{flippy} is available on GitHub~\cite{Dadunashvili_flippy_2023}, together with a full documentation and further demonstrations.
    The code used for the simulation in the \emph{Results} section is also part of \emph{flippy's} GitHub repository, and the most up-to-date version of it can be found in the \texttt{demo/biconcave\_shapes\_MC} folder of the repository~\cite{Dadunashvili_flippy_2023}.
    The version of the code that was most up to date at the time of writing this paper, and was used to generate the code snippets in the \emph{Results} section as well as the data for~\cref{fig:RBC_series}, can be found in reference~\cite{Dadunashvili_flippy_demo_2023}
    We use this code for several projects and have developed a robust workflow for introducing updates.
    New features are first used and tested by us before incorporating it into \emph{flippy}, which makes easier to detect and eliminate problems that slip by our unit testing framework.
    We intend to maintain and develop the code for years to come, while feedback and contributions from users are highly encouraged.

    \section*{Acknowledgments}
    We want to thank Emma Verhulst and Ian van Vliet for their use of early versions of flippy.
    Their feedback was instrumental during the refinement of \emph{flippy's} interface.
    We also want to thank Alexander Ziepke for his feedback on the manuscript of this paper.
    G.D. was supported by the “BaSyC – Building a Synthetic Cell” Gravitation grant (024.003.019) of the Netherlands Ministry of Education, Culture and Science (OCW) and the Netherlands Organisation for Scientific Research (NWO).\\

    \nolinenumbers
    \bibliography{main}

\end{document}